\documentclass[10pt,twocolumn,preprintnumbers,amsmath,amssymb,nofootinbib,superscriptaddress]{revtex4}
 
\usepackage{url}
\usepackage{hyperref}

\usepackage[normalem]{ulem}

\usepackage{latexsym}
\usepackage{epsfig}
\usepackage{amsmath}
\usepackage{amssymb}
\usepackage{wasysym}
\usepackage{graphicx}
\usepackage{verbatim}
\usepackage{enumerate,mdwlist}
\usepackage[titletoc]{appendix}
\usepackage{amsfonts}
\usepackage{pgfplots}
\usepackage[export]{adjustbox}
\usepackage{bbold}
\usepackage[normalem]{ulem}
\usepackage{fixdif}
\usepackage{acro}
\usepackage{graphicx} 

\DeclareAcronym{LVK}{short=LVK, long=LIGO-Virgo-KAGRA}
\DeclareAcronym{O5}{short=O5, long=the fifth LVK observing run}
\DeclareAcronym{O3}{short=O3, long=the third LVK observing run}
\DeclareAcronym{O4}{short=O4, long=the fourth LVK observing run}
\DeclareAcronym{GWTC-3}{short=GWTC-3, long=the third gravitational-wave transient catalog}
\DeclareAcronym{GW}{short=GW, long=gravitational wave}
\DeclareAcronym{CBC}{short=CBC, long=compact binary coalescence}
\DeclareAcronym{BBH}{short=BBH, long=binary black hole}
\DeclareAcronym{BNS}{short=BNS, long=binary neutron star}
\DeclareAcronym{SNR}{short=SNR, long=signal-to-noise ratio}
\DeclareAcronym{FAR}{short=FAR, long=false alarm rate}
\DeclareAcronym{PISN}{short=PISN, long=pair-instability supernova}
\DeclareAcronym{PPD}{short=PPD, long=posterior predictive distribution}

\newcommand{\ba}{\begin{eqnarray}}
\newcommand{\ea}{\end{eqnarray}}
\newcommand{\be}{\begin{equation}}
\newcommand{\ee}{\end{equation}}

\newcommand{\dd}{\mathrm{d}}
\newcommand{\yr}{\mathrm{yr}}

\newcommand{\mmax}{m_\mathrm{max}}

\newcommand{\Gpc}{\mathrm{Gpc}}

\newcommand{\Msun}{M_\odot}

\newcommand{\tgf}{\tau^{\geq 5}}

\definecolor{grey}{rgb}{0.4,0.4,0.4}
\definecolor{dullmagenta}{rgb}{0.4,0,0.4}
\definecolor{darkblue}{rgb}{0,0,0.4}
\definecolor{midblue}{rgb}{0,0,0.5}
\definecolor{midred}{rgb}{0.5,0,0}
\definecolor{orange}{rgb}{1,0.5,0}
\definecolor{lightbrown}{rgb}{0.75,0.5,0.25}
\definecolor{tan}{cmyk}{0.14,0.42,0.56,0}
\definecolor{djunglegreen}{cmyk}{0.99,0,0.52,0}
\definecolor{lightgreen}{rgb}{0,1,0}
\definecolor{olivegreen}{cmyk}{0.64,0,0.95,0.40}
\definecolor{midgreen}{rgb}{0.0,0.675,0.0}
\definecolor{darkgreen}{rgb}{0,0.5,0}

\begin{document}

\title{Gravitational lensing rarely produces high-mass outliers to the compact binary population 
}

\author{Amanda M. Farah}
\email{afarah@uchicago.edu}
\affiliation{Department of Physics, University of Chicago, Chicago, IL 60637, USA}

\author{Jose Mar\'ia Ezquiaga}
\affiliation{Center of Gravity, Niels Bohr Institute, Blegdamsvej 17, 2100 Copenhagen, Denmark}

\author{Maya Fishbach}
\affiliation{Canadian Institute for Theoretical Astrophysics, David A. Dunlap Department of
Astronomy and Astrophysics, and Department of Physics, 60 St George St, University of Toronto, Toronto, ON M5S 3H8, Canada}

\author{Daniel E. Holz}
\affiliation{Department of Physics, University of Chicago, Chicago, IL 60637, USA}

\begin{abstract}
All gravitational-wave signals are inevitably gravitationally lensed by intervening matter as they propagate through the Universe.
When a gravitational-wave signal is magnified, it \emph{appears} to have originated from a closer, more massive system.
Thus, high-mass outliers to the gravitational-wave source population are often proposed as natural candidates for strongly lensed events.
However, when using a data-driven method for identifying population outliers, we find that high-mass outliers are not necessarily strongly lensed, nor will the majority of strongly-lensed signals appear as high-mass outliers.
This is both because statistical fluctuations produce a larger effect on observed binary parameters than does lensing magnification, and because lensing-induced outliers must originate from intrinsically high-mass sources, which are rare. 
Thus, the appearance of a single lensing-induced outlier implies the existence of many other lensed events within the catalog.
We additionally show that it is possible to constrain the strong lensing optical depth, which is a fundamental quantity of our Universe, with the detection or absence of high-mass outliers.
However, constraints using the latest gravitational-wave catalog are weak---we obtain an upper limit on the optical depth of sources at redshift $1$ magnified by a factor of $5$ or more of $\tau(\mu\geq5,z=1)\leq$0.035---and future observing runs will not make an outlier-based method competitive with other probes of the optical depth.
However, the full inferred population of compact binaries may be more informative of the distribution of lenses in the Universe, opening a unique opportunity to access the high-redshift Universe and constrain cosmic structures.
\end{abstract}

\date{\today}

\maketitle

\section{Introduction}
\begin{figure*}
    \centering
    \includegraphics[width=\textwidth]{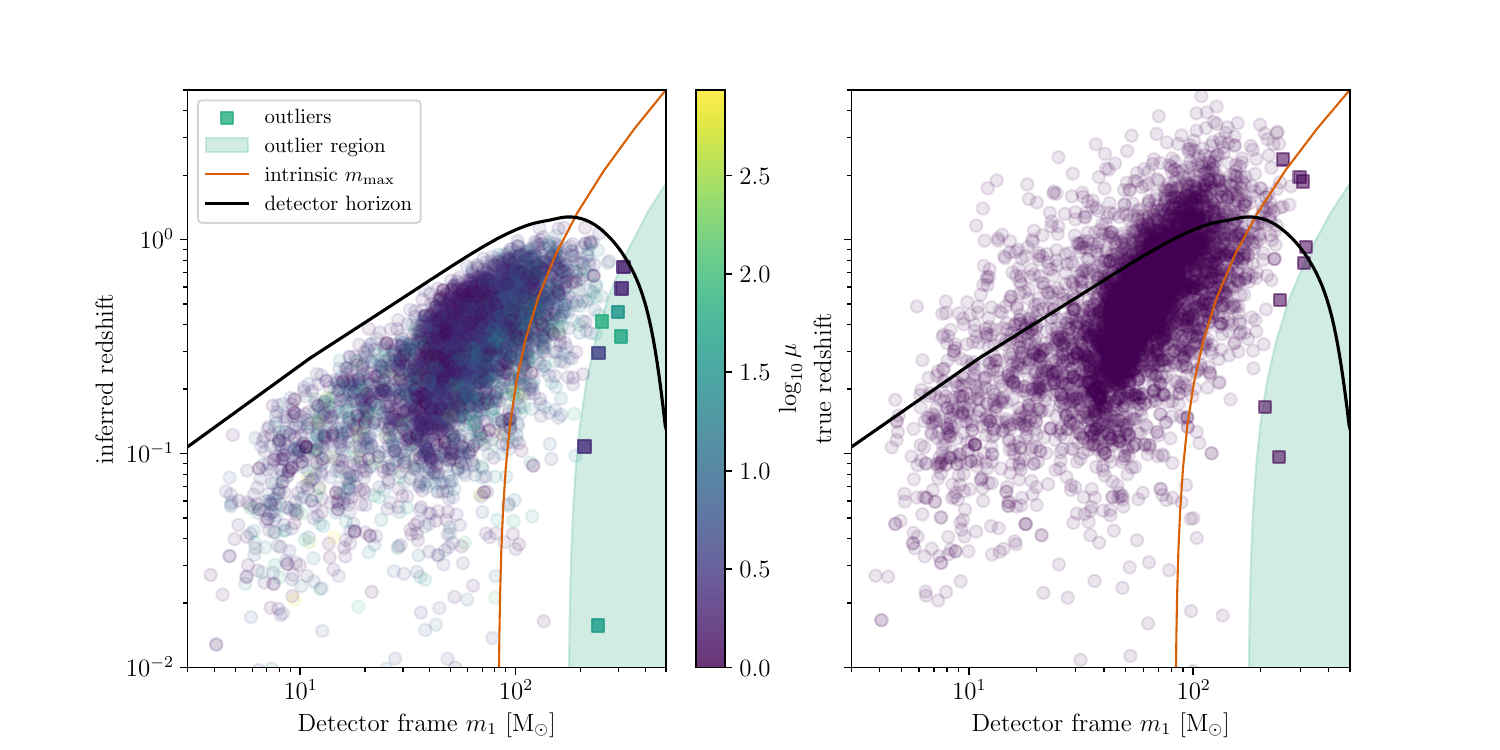}
    \caption{Simulated observed BBH population with an \acs{O3}-like detector with (\emph{left panel}) and without (\emph{right panel}) the effects of gravitational lensing. To clearly demonstrate these effects, the magnification distribution used for this plot has unrealistically high support for large magnifications. The same systems are plotted in both panels. In both panels, we show the detector horizon (\emph{black line}), the maximum true source frame mass (\emph{orange line}), and the region in which we define events to be outliers, as described in Section~\ref{sec:methods-ffh} (\emph{green shading}). The separation between the maximum mass of the true population and the outlier region is due to the large measurement uncertainties typical of GW observations. In the left panel, binaries are colored by their lens magnification. Magnification causes the binaries' apparent redshifts ($y$-axis of left panel) to be lower than their true redshifts ($y$-axis of right panel). This allows for the detection of events whose true distances lie beyond the detector horizon and results in larger apparent source frame masses. Thus, most high-mass outliers (\textit{squares}) have large magnifications. However, events with large true redshifts and source frame masses are more likely to become outliers than low-mass events, as indicated by the cluster of squares in the upper-right corner of the right plot. ``False positives'' are also possible, as evidenced by the two events whose true parameters lie in the outlier region.}
    \label{fig:schematic}
\end{figure*}

\Acp{GW} from \acp{CBC} are unique probes of cosmic structures. 
These binaries merge at cosmological distances and their emitted \ac{GW} signals are understood from first principles.
As they travel through the Universe, \acp{GW} are not absorbed by intervening matter and are therefore only altered by gravitational lensing~\cite{xu_please_2022}. 
Additionally, the selection biases of \ac{GW} detectors is known precisely as a function of \ac{CBC} parameters.
Thus, \acp{GW} are uniquely clean probes of gravitational lensing and compliment lensed electromagnetic transients~\cite{oguri_strong_2019,liao_strongly_2022}---such as supernovae~\cite{suyu_strong_2024}, gamma-ray bursts~\cite{levan_gravitational_2025}, fast radio bursts~\cite{pastor-marazuela_fast_2025}, and quasars~\cite{jackson_quasar_2013}---whose selection and propagation effects are more complex.

Gravitational lensing affects all sources (weak lensing), although only a small fraction are significantly magnified (strong lensing). 
When high magnifications occur, gravitational lensing provides an opportunity to observe the properties of systems that lie beyond the detectors' nominal horizon~\cite{oguri_gravitational_2003, rydberg_detecting_2020}. 
Because of \ac{GW} detectors' millisecond-level time resolution~\cite{acernese_advanced_2014,aasi_advanced_2015,akutsu_overview_2021}, \acp{GW} observations are sensitive to a wide mass range of lenses, from stellar mass compact objects to galaxy clusters.
For large lenses, strong lensing of \acp{GW} appears as repeated signals of the original merger, while for small lenses, the \ac{GW} waveform could be distorted due to interference and diffraction \cite{deguchi_diffraction_1986,nakamura_wave_1999}. 
Although current searches have not found evidence for lensing so far \citep{hannuksela_search_2019,abbott_search_2021,janquart_follow-up_2023,abbott_search_2024}, a strongly-lensed signal will inevitably be present in \ac{GW} catalogs as they grow to include thousands of events~\cite{ng_precise_2018,li_gravitational_2018,xu_please_2022}.
Identifying such events will be crucial to avoid bias in population inference and searches for modified gravity~\cite{dai_effect_2017,oguri_effect_2018,ezquiaga_modified_2022,vijaykumar_detection_2023,canevarolo_impact_2024}.

In this work, we explore the magnification caused by strong gravitational lensing: the lens causes a convergence of rays, resulting in an image with more power than it would otherwise have without a lens.
For \acp{GW}, a magnification $\mu$ is observed as a change in amplitude of the gravitational waveform: $h^l = \sqrt{\mu} h$, where $\mu$ is the traditional magnification factor, $h$ is the intrinsic strain, and $h^l$ is the strain of the lensed signal.
Additionally, for \acp{GW} from \acp{CBC}, the unlensed amplitude is inversely proportional to the luminosity distance, $d_L$, to the source.
Thus, a magnified \ac{CBC} source will appear closer than an unlensed source.
The true luminosity distance $d_L^t$ of a lensed \ac{CBC} is related to its apparent luminosity distance $d_L^l$ by
\begin{equation}
    d_L^t=\sqrt{\mu}\, d_L^l \, .
    \label{eq:true vs apparent dl}
\end{equation}
Herein, we define \emph{true} parameters to be those intrinsic to the \ac{CBC}.
We define \emph{apparent} parameters to be those that would be measured from a \ac{GW} signal in the absence of detector noise, differing from the true parameters only by the effects of magnification.
Finally, \emph{observed} parameters are those that are measured from the combination of the \ac{GW} signal and detector noise, and therefore include the effects of measurement error and lensing. 

Eq.~(\ref{eq:true vs apparent dl}) showcases a perfect degeneracy between the magnification of the source and its apparent luminosity distance, making it impossible to determine the magnification of a signal without knowing its luminosity distance, and vice versa.
Magnification also affects the apparent masses of \acp{CBC}.
Masses are measured in the detector frame,  and are related to their intrinsic, or source-frame quantities through the redshift to the source.
However, redshift is not directly measured from the \ac{GW} signal; it is calculated based on the signal's measured luminosity distance and an assumed cosmological model.
Thus, inferring an erroneously small luminosity distance causes the inference of an erroneously small redshift and therefore an erroneously large source-frame mass.

Exceptionally high- or low-mass systems are therefore promising candidates for gravitationally-lensed \ac{GW} signals.
This is demonstrated in Fig.~\ref{fig:schematic}, which shows the redshifts and detector-frame masses of simulated \acp{BBH}.
True \ac{BBH} parameters are drawn from a known astrophysical distribution, lensed by an exaggerated lens model, and a fraction of them are detected based on their observed parameters (left panel).
Only the detected systems are plotted, and the same systems are shown on both panels; the only difference is that the left panel plots apparent redshifts and the right panel plots unlensed redshifts.
The right panel includes the effects of measurement uncertainty but not the effects of lensing.
The population from which the systems are drawn has a high-mass truncation, $\mmax$, beyond which no true source-frame masses are possible.
This is converted to a detector-frame mass and plotted as an orange line in both panels.
However, measurement uncertainty causes a discrepancy between true and observed parameters, so some events' masses are observed to be higher than $\mmax$ even without the effects of magnification.
This is demonstrated by the fact that some points lie to the right of the orange line in Fig.~\ref{fig:schematic}'s right panel.
Our definition of an exceptionally high-mass event must therefore take measurement uncertainty into account~\citep{fishbach_most_2020}.
Doing so results in the green shaded region in both panels.
Systems whose observed parameters lie in this outlier region are indicated by squares.

For the \ac{CBC} population and lens model considered in Fig.~\ref{fig:schematic}, the majority of detected population outliers are highly magnified events.
However, most detected outliers have large true masses: the squares on the right panel lie near the orange line. 
This means that the effect of lensing on the observed source frame mass is small compared to the dynamic range of the \ac{BBH} mass distribution.
It follows that many binaries with large magnifications do not become high-mass outliers, as their true masses and redshifts are not large enough.
Because high-mass systems are rare compared to low-mass ones \citep{abbott_binary_2019,abbott_population_2021,abbott_population_2023}, the existence of even one high-mass outlier due to lensing implies that many other lower-mass events should have been lensed with comparable magnifications. 
``False positives'' are also possible: a large noise fluctuation can make an unlensed system's observed parameters fall within the outlier region, as shown by the event in the green region of Fig.~\ref{fig:schematic}'s right panel.

Regardless, Fig.~\ref{fig:schematic} shows that high-mass events are candidates for gravitational lensing.\footnote{Exceptionally low-mass events may also be promising candidates for de-magnification ($\mu<1$), but their decreased amplitude makes them less likely to detect, so we do not search for low-mass outliers in this work.}
This has previously been pointed out in \citet{bianconi_gravitational_2023,smith_discovering_2023}, and \citet{janquart_what_2024}
in the context of lensed \acp{BNS} and is a hypothesis explored in \citet{abbott_properties_2020}, because of the exceptionally high mass of the \ac{BBH} GW190521. 
\citet{broadhurst_reinterpreting_2018,broadhurst_uniform_2022} have even argued that all \ac{GW} events with observed chirp masses above $\sim20\Msun$ are gravitationally lensed. 
These studies all assume an exact form for the \ac{CBC} mass distribution and interpret observed masses above this assumed (true mass) limit as having been caused by lens magnification.
However, it is statistically inconsistent to compare observed and true parameters, as the latter does not take measurement uncertainty into account. 
Furthermore, while the assumed form of the mass distribution differs between these studies, none of them are  informed by the \ac{GW} data. 
\citet{broadhurst_reinterpreting_2018,broadhurst_uniform_2022} use electromagnetic observations of galactic X-ray binaries to justify an upper limit to the \ac{BBH} mass distribution \citep{remillard_x-ray_2006}, \citet{smith_discovering_2023} also use X-ray binary observations to motivate the assumption of a perfectly empty lower mass gap between \acp{BNS} and \acp{BBH} \citep{ozel_black_2010,farr_mass_2011}, and \citet{abbott_properties_2020} assume a maximum black hole mass informed by theoretical expectations for a pair instability supernova feature.
While these population features are either physically or observationally motivated, they may not apply to the population of \ac{GW} sources. 
In fact, these features are not necessitated by existing \ac{GW} data \citep{fishbach_does_2020, fishbach_apples_2022,farah_bridging_2022, abbott_population_2023} if lensing is rare enough to affect only $\mathcal{O}(1)$ event in current catalogs.
This motivates us to explore the prospects for identifying lensed \ac{GW} signals through their statistical inconsistency with the \ac{CBC} population as inferred from \acp{GW} in a self-consistent and statistically rigorous manner. 

Because gravitational lensing causes an increased number of high-mass outliers, a lack of outlier detections allows us to place upper limits on the rate of gravitational lensing.
This rate can be parametrized by the high-magnification optical depth, a fundamental quantity describing the properties of our Universe. 
The optical depth encodes invaluable information about the distribution of lenses and their ability to produce caustics that could highly magnify a signal~\cite{blandford_fermats_1986, holz_new_1998,holz_gravitational_1999}.
It therefore informs the history of structure formation and the nature of dark matter.
In this paper, we place an upper limit on the high-magnification optical depth using current population measurements and the lack of high-mass population outliers ~\cite{abbott_population_2023} in \ac{GWTC-3}~\citep{collaboration_gwtc-3_2021}.

This paper is organized as follows.
In Section~\ref{sec:methods}, we summarize our current understanding of the \ac{BBH} population from GW observations and review the method for identifying population outliers that was originally developed in \citet{fishbach_most_2020}.
We also introduce a parametrization for the distribution of lens magnifications that we use in this work and derive the statistical framework necessary to constrain it. 
In Section~\ref{sec:results-gwtc3}, we present constraints on the magnification distribution using \ac{GWTC-3} data.
Section~\ref{sec:results-future} provides projections for future detectors' constraints on the magnification in the case of continued non-detection of outliers as well as if a single high-mass outlier is identified. 
We conclude and discuss future work in Section~\ref{sec:discussion}.
The code used to produce all figures and numbers in this paper is made public at \url{github.com/afarah18/GW-lensing-outliers-public} \cite{farah_afarah18gw-lensing-outliers-public_2025}.

\section{Methods}
\label{sec:methods}
In this section we introduce our parametrization of the magnification distribution as well as our assumed astrophysical population of \acp{CBC}.
We then summarize the statistical framework developed in \citet{fishbach_most_2020} to determine if a given \ac{GW} event is a population outlier while accounting for measurement uncertainty, population model uncertainty, and detector noise. 
Finally, we demonstrate how we use the (non)detection of population outliers to constrain the lensing optical depth.

\subsection{Lens magnification model, $p(\mu|z)$}
\label{sec:methods-mu distribution}
\begin{figure}
    \centering
    \includegraphics[width=\linewidth]{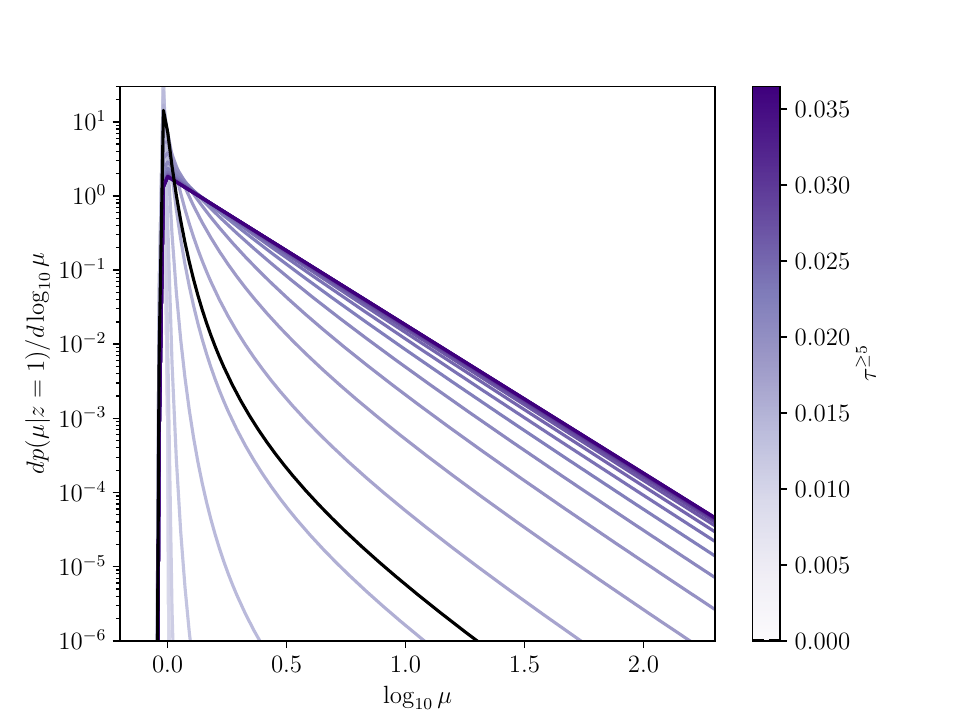}
    \caption{Magnification distribution at $z=1$ for various choices of the tail width parameter, $t^0_c$. 
    These directly correspond to different optical depths, which are denoted by the colorbar. 
    The black line indicates our fiducial model, which is the one presented in \protect{\citet{dai_effect_2017}} and calibrated to cosmological simulations. 
    It corresponds to $\tgf=1.47\times 10^{-5}$ (\emph{black dashed line on colorbar}).}
    \label{fig:p mu tau}
\end{figure}
The number of high-mass outliers is sensitive to the redshift-dependent distribution of lens magnifications, $p(\mu|z)$ and, equivalently, the optical depth of strong gravitational lensing, $\tau$~\cite{holz_seeing_2001}. 
We must parametrize $p(\mu|z)$ so that we may constrain the parameters governing its morphology using the (non-)observation of lensed \acp{GW}.
We use the parametrization presented in \citet{dai_effect_2017} as our fiducial model
(though using a different lens model does not significantly impact our results; see Appendix~\ref{ap:robustness checks:p_mu}):
\begin{equation}
\begin{aligned}
    \frac{\dd P(\mu)}{\dd \ln \mu} = A(t_0)\int_0^\infty  \frac{\dd t}{\sqrt{2\pi}\sigma} \exp&\Bigg[\frac{5}{t+t_0} - 2t \\
    & - \frac{(\ln\mu - \delta - t)^2}{2\sigma^2}\Bigg] ,
    \label{eq:Dai17}
\end{aligned}
\end{equation}
where $A$ is a normalization factor,
\begin{equation}
    A(t_0) = \left[\int_0^\infty \dd t \exp\left[\frac{5}{t+t0}-2t\right]\right]^{-1} .
\end{equation}
Here, $\sigma$ and $\delta$ characterize the width and mean of the log-normal distribution, and $t_0$ controls the width of the heavy-tailed kernel with which the log-normal distribution is convolved.
Each of these three parameters depends on redshift.
Their redshift dependence is chosen such that Eq.~(\ref{eq:Dai17}) fits the results from cosmological simulations at various redshifts.
The results of these calibrations are reported in Table 1 of \citet{dai_effect_2017}, which we also adopt for the available redshift range, $0.7<z<20$.  
We validate this phenomenological model with the halofit prescription of \cite{takahashi_revising_2012}, which was used in \cite{oguri_effect_2018}. 
This also allows us to calibrate $\sigma$, $\delta$, and $t_0$ in the range $0<z<0.7$.

Our method is primarily sensitive to the width of the high-magnification tail, as parametrized by $t_0$.
We add an additional parameter, $t_0^c$, to the model described in Eq.~(\ref{eq:Dai17}).
This parameter is an additive constant to $t_0$, so when  $t_0^c=0$, we have the fiducial model in Eq.~(\ref{eq:Dai17}).
This maintains the tail width's redshift dependence while allowing it to increase at fixed redshift by increasing $t_0^c$.
The effects of different choices of $t_0^c$ on the magnification distribution are shown in Fig.~\ref{fig:p mu tau}.

Neither $t_0^c$ nor any of the parameters in our fiducial model have clear physical interpretations.
Therefore, rather than reporting constraints on $t_0^c$ or any other parameter, we will always report the $\tau$ induced by our constraints on $t_0^c$.
Given a form for the magnification distribution, we can calculate $\tau$ at any reference redshift $z^{\rm ref}$ and above any $\mu^{\rm ref}$ via 
\begin{equation}
    \tau^{\geq \mu^{\rm ref}}(z^{\rm ref}) = \int_{\mu^{\rm ref}}^{\infty}\dd \mu p(\mu|z=z^{\rm ref}) .
\end{equation}
For concreteness, we calculate $\tau$ at $z=1, \mu\geq5$ for the remainder of this work and will refer to this derived parameter as $\tgf$.
This choice is arbitrary, and was made for easier comparison to the $z=1$ strong-lensing optical depth often referenced in strong-lensing literature.

\subsection{Astrophysical populations of compact binaries}
\label{sec:methods-astro pop}
To define population outliers and simulate GW events, we must assume a population model for the masses and distances of GW sources. 
We focus our main analysis on the population of \acp{BBH}, but also comment on \acp{BNS}.

\subsubsection{Binary black holes}
Our distribution of primary masses and mass ratios takes the form of the \textsc{Power Law + Peak} model from \citet{talbot_measuring_2018,abbott_population_2023} for the distribution of primary masses and mass ratios of \acp{BBH}.\footnote{During the review of this Article, GWTC-4 \cite{abac_gwtc-40_2025} was released, along with updated population models. Adopting an entirely different population model will affect the expected number of high-mass outliers. However, in the region of interest to this study, \textsc{Power Law + Peak} under a fit to GWTC-3 is very similar to the updated population model under a fit to GWTC-4 (see the high-mass portion of Fig. 3 of \citet{abac_gwtc-40_2025-1}), so we do not expect a substantive change to our conclusions under this new model.}
We use the redshift distribution from \citet{callister_shouts_2020}, which is consistent with that inferred in \citet{abbott_population_2021}, but extends to higher redshifts. 
These two distributions are independent; i.e. the mass distribution does not evolve with redshift.
This assumption is likely accurate: \citet{fishbach_when_2021,van_son_redshift_2022,abbott_population_2023,lalleman_no_2025} have analyzed the current \ac{BBH} observations and found that if the mass distribution evolves with redshift, this evolution must be fairly small in the ranges accessible to current detectors.
As an aside, the lack of evidence for a redshift-evolving mass distribution also constrains the magnification distribution, as strong lensing would cause higher mass black holes at higher redshifts, as demonstrated by \citet{dai_effect_2017, oguri_effect_2018, broadhurst_uniform_2022}.

The parameters we use to evaluate our population model differ for different parts of the analysis.
As we discuss in Sections~\ref{sec:methods-ffh} and \ref{sec:methods-Nexp}, we marginalize over the full hyperposterior when constructing our definition of an outlier and fix population parameters to the hyperposterior mean when generating injections with which to evaluate the expected number of outliers.

We do not explicitly model the spin distribution throughout this work as they have a negligible effect on the detectability of \ac{BBH} mergers.
Additionally, spin inference is not impacted by gravitational lensing magnification, though it is impacted by other lensing effects \citep[e.g.][]{mishra_exploring_2024}.
For the purposes of estimating selection effects, we assume all injections have zero spin magnitudes.

\subsubsection{Binary neutron stars}
To facilitate comparison with \citet{smith_discovering_2023}, we adopt a uniform distribution of \ac{BNS} primary masses between $1$ and $2.5\Msun$.
We additionally adopt their redshift distribution, which has the same form as that of \citet{callister_shouts_2020}, but with a lower peak redshift of $1.9$.
We do not consider the effect of spins.

\subsection{Identifying population outliers}
\label{sec:methods-ffh}

To identify population outliers, we follow the procedure presented in \citet{fishbach_most_2020} to define a threshold mass $\mmax^{\rm thresh}$ above which observations are deemed inconsistent with the population inferred from the rest of the data.
In particular, we first construct a \ac{PPD} of the maximum observed mass (accounting for both measurement uncertainty and selection effects) out of $N$ detected events, where $N$ is the number of observations in a given catalog.
We then define outliers as events with observed primary mass larger than the $P$th percentile of this \ac{PPD}.
As we show in Appendix~\ref{ap:robustness checks:P_thresh}, the value of $P$ is somewhat arbitrary, and lowering $P$ increases the number of false positives, with low values of $P$ resulting in more false positives than true positives.
We therefore attempt to choose a value that minimizes the ratio of the expected number of outliers in a Universe with no lensing (i.e. the false positive rate), to the expected number of outliers in a Universe with lensing (the sum of the false positive and true positive rates). 
We simulated several detected populations and found that, when the optical depth is large, the ratio of these two rates is lowest when $P\geq99$. 
We therefore use $P=99$ for the remainder of this work. 
Under this choice, the ratio of false positives to the sum of false and true positives is $\approx0.015$ for magnification distributions with large optical depths, $\approx0.15$ for those with moderate ones, and near unity for our fiducial magnification model.

To sample from the \ac{PPD}, we use the \texttt{GWMockCat} code \citep{farah_things_2023} to simulate $N$ observed events with the detector sensitivity and measurement uncertainty typical of the catalog in question.
For example, for \ac{GWTC-3}, $N=69$, as we define our catalog as the events used for the BBH population fits in \citet{abbott_population_2023}.
We then take the maximum-likelihood value of the primary masses of each event, and find the largest value.
This is one draw from the \ac{PPD}.
We build the \ac{PPD} by repeating this procedure $\approx 2,000$ times, marginalizing over the uncertainty in the population hyperparameters inferred by \citet{abbott_population_2023}.
When we consider future catalogs, we scale the hyperparameter uncertainty by $1/\sqrt{N}$ to account for the decrease in statistical uncertainty caused by additional events.

A key assumption made in this work is that the population used to construct the \ac{PPD} is the population of unlensed \acp{CBC}, and outliers are defined with respect to this unlensed population.
This is equivalent to assuming that we infer the \ac{CBC} population at $z\sim0$ and assume that it does not evolve astrophysically within our sensitive volume.
We addressed the latter part of this assumption in Section~\ref{sec:methods-astro pop}.
However, the assumption that we are able to infer the unlensed distribution is less likely with high-$\tau$ models.
Future work will eliminate the need for this assumption by simultaneously fitting the magnification distribution with the \ac{CBC} population.

\begin{figure*}
    \centering
    \includegraphics[width=\linewidth]{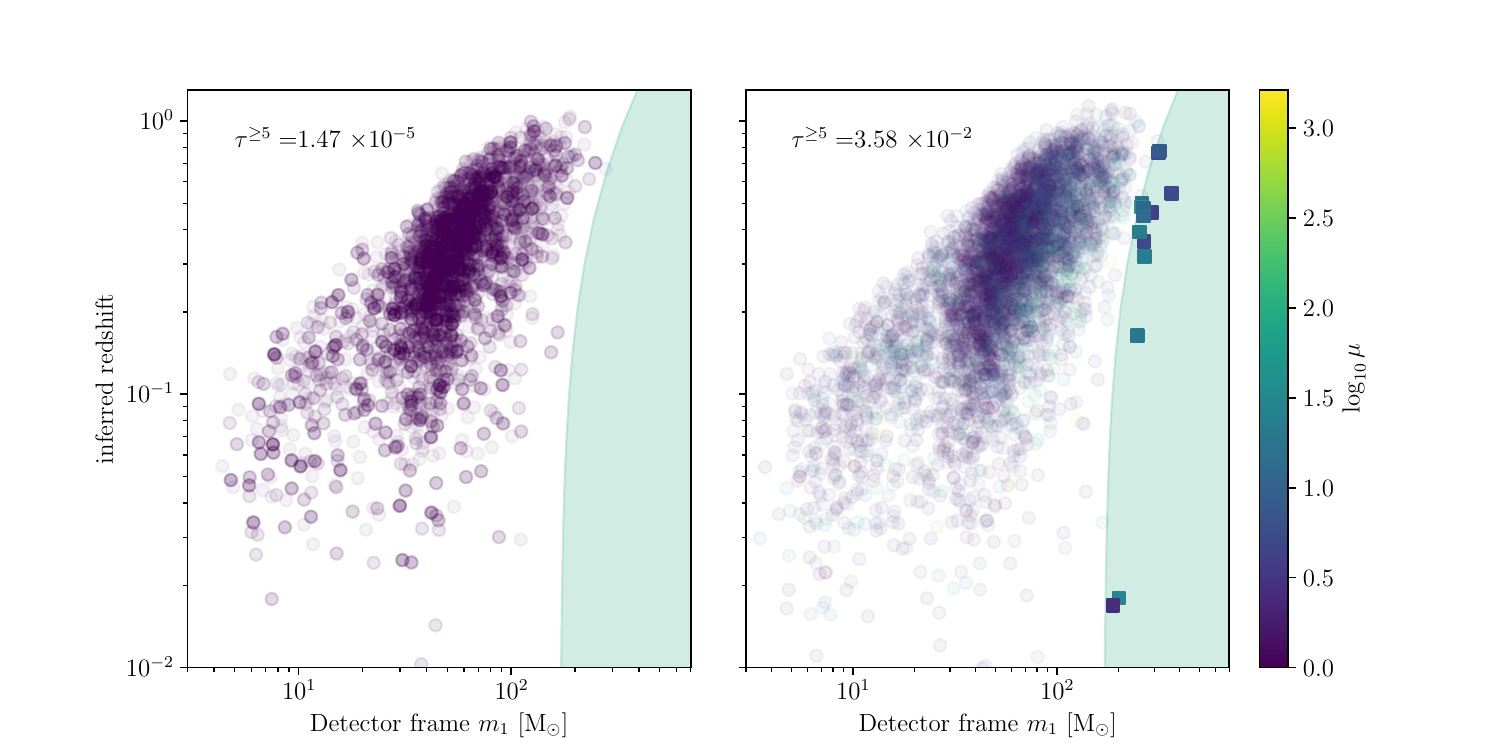}
    \caption{Similar to the left panel of Fig.~\ref{fig:schematic}, but with two different values of the high-magnification optical depth, $\tgf$, and an expanded colorbar range. Larger values of $\tgf$ allow for a higher fraction of events to receive large magnifications, causing more observed outliers. This allows the number of outliers to constrain the width of the high-magnification tail. Additionally, the observation of even one high-mass outlier implies the existence of many lower-mass events with comparable or higher magnifications.}
    \label{fig:tau and outliers}
\end{figure*}

\subsection{Statistical framework}
\label{sec:methods-Nexp}
Now that we have defined a method to classify and count population outliers, we can use the number of outliers to constrain the lensing optical depth.
The detection of population outliers is a Poisson process. The probability of observing $k$ outliers is therefore given by
\begin{align}
p(k|N_{\exp}^{\rm outlier}) = \frac{e^{-N_{\exp}^{\rm outlier}}}{k!}\left(N_{\exp}^{\rm outlier}\right)^k ,
\label{eq:poisson}
\end{align}
where $N_{\exp}^{\rm outlier}$ is the number of expected outliers. This is given by
\begin{equation}
\begin{aligned}
    N_{\text{exp}}^{\rm outlier} = \int \dd \mu& \dd m_1 \dd m_2 \dd z \dd t \dd \hat{n} \Bigg[ \frac{\dd N}{\dd z \dd t \dd m_1 \dd q \dd \mu} \\
    & \times P_{\det + \mathrm{outlier}}(m_1, q, z, \mu, \hat{n})\Bigg],
    \label{eq:Nexp general}
\end{aligned}
\end{equation}
where $ P_{\det + \mathrm{outlier}}(m_1, q, z, \mu, \hat{n})$ is the probability that a source with true parameters $m_1, q, z,$  and $\mu$ and noise realization $\hat{n}$ is both detected and identified as a high-mass outlier.
Hence, $P_{\det + \mathrm{outlier}}(m_1, q, z, \mu, \hat{n})$ accounts for both measurement uncertainty and selection effects.
$\dd N/\dd z \dd t \dd m_1 \dd q \dd \mu$ is the number of mergers per unit redshift, time, primary mass, mass ratio, and magnification.
This is also known as the differential merger rate or the population model.
Our assumed form for $\dd N/\dd z \dd t \dd m_1 \dd q \dd \mu$ is given in Sections~\ref{sec:methods-mu distribution} and ~\ref{sec:methods-astro pop}.
In particular, we assume that the lens magnification distribution is independent of the population of binary black hole mergers.
This is equivalent to assuming that the size and abundance of dark matter halos is disconnected from the population of BBH mergers. 
Such assumption would be broken in the high-magnification regime where diffraction, which is determined by the \ac{GW} wavelength, sets the maximum magnification \cite{lo_observational_2025,maria_ezquiaga_diffraction_2025}.
For signals in the \ac{LVK} band and galaxy-scale or cluster-scale lenses, this corresponds to $\mu\sim \mathrm{few} \times 1000$.
Under the assumption of independent magnification and BBH distributions, we can write Eq.~(\ref{eq:Nexp general}) as
\begin{equation}
\begin{aligned}
    N_{\text{exp}}^{\rm outlier} =& \int_0^\infty \dd \mu \int \dd m_1 \dd m_2 \dd z \dd t \dd\hat{n}\Bigg[\frac{\dd N}{\dd z\dd t} p(m_1,q) \\
    &\times p(\mu|z,t^c_0) P_{\det + \mathrm{outlier}}(m_1, q, z, \mu, \hat{n})\Bigg].
    \label{eq:Nexp specific}
\end{aligned}
\end{equation}
Here, $\dd N/\dd z \dd t$ is the number of mergers per unit redshift per unit time (the volumetric rate), $p(m_1,q)$ is the two-dimensional mass distribution of compact mergers, and $p(\mu|z)$ is the magnification distribution.

In order to calculate $N_\mathrm{exp}^\mathrm{outlier}$ as a function of the lensing magnification distribution, we use the mass and redshift distributions described in Section~\ref{sec:methods-astro pop}, assume a local BBH merger rate of $15 \, \Gpc^{-3} \yr^{-1}$, and fix the parameters of the redshift and mass models to the mean values of the hyperposterior obtained through our fit to GWTC-3 data, as well as those inferred in \citet{callister_shouts_2020}.
In principle, different choices for these population parameters will impact the expected number of observed outliers, because a population that creates more high-mass events results in more outliers and a higher predicted probability of seeing an outlier.
However, as we show in Appendix~\ref{ap:robustness checks:Lambda}, varying these hyperparameters within the values allowed by their posterior does not dramatically change our conclusions.
We use the magnification distribution described in Section~\ref{sec:methods-mu distribution}.

We calculate $N_{\exp}^{\rm outlier}$ by estimating Eq.~(\ref{eq:Nexp specific}) as a Monte-Carlo sum:
\begin{equation}
    N_{\text{exp}}^{\rm outlier} \approx \sum_{\mathrm{outliers}, i} \left.\frac{\dd N}{\dd z\dd t} \right|_{z^i,t^i}p(m_1^i,q^i) p(\mu^i|z^i,t^0_c) .
    \label{eq:Nexp MC}
\end{equation}
Algorithmically, we evaluate Eq.~(\ref{eq:Nexp MC}) by drawing GW event parameters from the above-described population of masses, redshifts, and magnifications. 
We then use the semi-analytic injection generation scheme in the \texttt{GWMockCat} code \citep{farah_things_2023} to determine which events would be observed with a given detector's sensitivity and its expected measurement uncertainty, marginalizing over noise realizations $\hat{n}$.
Projected measurement uncertainties from future detectors are taken from Table 1 of \citet{ezquiaga_jumping_2021} and measurement uncertainties typical of \ac{O3} are taken from \citet{farah_things_2023}.
Current and future detector sensitivities are taken from \citet{abbott_prospects_2020}.
Next, we determine which events would be identified as outliers using the procedure in Section~\ref{sec:methods-ffh}.
Finally, we evaluate the expression in Eq.~(\ref{eq:Nexp MC}), summing over the identified and found outliers. 
This yields the expected number of outliers as a function of $t^0_c$, $N_{\exp}^{\rm outlier}(t^0_c)$.
As discussed in Section~\ref{sec:methods-mu distribution}, this is straightforwardly transformed into a function of the optical depth at a specific redshift and magnification.

Given $N_{\exp}^{\rm outlier}(\tgf)$ and Eq.~(\ref{eq:poisson}), we can calculate the probability of observing $k$ outliers as a function of $\tgf$.
If no outliers are observed, the posterior on $\tgf$ takes the form of a decaying exponential in $N_{\exp}^{\rm outlier}$, and we can only provide an upper limit on $\tgf$.
If a high-mass population outlier is detected in the future and it can be attributed to gravitational lensing, we can set $k=1$ to make a measurement on $\tau$.

Eqs.~(\ref{eq:Nexp specific}) and (\ref{eq:Nexp MC}) make it clear that $ N_{\text{exp}}^{\rm outlier}$ depends on the parameters of the magnification distribution, of which we only vary $t^0_c$.
The effect of the magnification distribution on the number of expected outliers can also be seen in Fig.~\ref{fig:tau and outliers}, where we show the detected population of GW events for two magnification distributions with different optical depths.
Our fiducial population, which has a low optical depth, results in 0 outliers out of $5,000$ detected events, whereas the population with a large optical depth results in 14 outliers.
The increase in high-mass outliers due to large optical depths is the basis of this work. 

\section{Constraints on the optical depth with GWTC-3}
\label{sec:results-gwtc3}
\begin{figure}
    \centering
    \includegraphics[width=\linewidth]{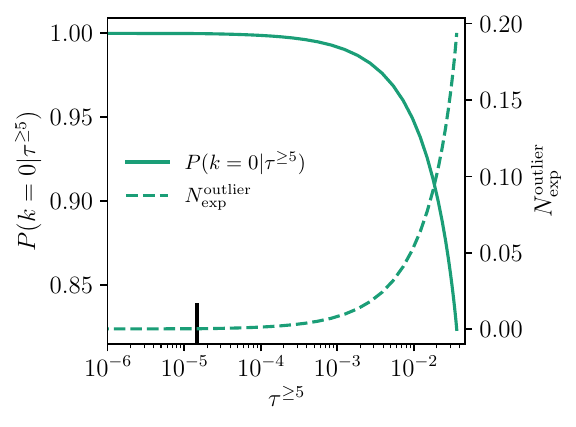}
    \caption{Expected number of outliers in GWTC-3 (\emph{dashed line, right $y$-axis}) and the resulting probability of observing $k=0$ outliers in GWTC-3 (\emph{solid line, left $y$-axis}) as functions of the high-magnification optical depth, $\tgf$.
    Though the expected number of outliers increases monotonically with the optical depth, it is still less than unity even for extreme values of $\tgf$.
    For all allowed $\tgf$, the probability of not having observed an outlier is $\gtrsim 85\%$, meaning that meaningful constraints on the optical depth are not possible with current \ac{GW} data.
    Our fiducial value for $\tgf$ is indicated by the black tickmark.
    }
    \label{fig:tau gwtc-3}
\end{figure}

Applying our outlier definition to GWTC-3,  we find no high-mass outlier detections, consistent with previous work~\citep{abbott_population_2021,abbott_population_2023,abbott_properties_2020}.
We reiterate that the population model assumed for this outlier definition is based on a fit to the GWTC-3 data, under the assumption that no events in the catalog (except for the outlier candidate) are lensed.
The lack of high-mass outlier detections in GWTC-3 translates to an upper limit on the optical depth, $\tau$.
In Fig.~\ref{fig:tau gwtc-3}, we show the probability of observing zero outliers with an \ac{O3}-like LIGO detector as a function of $\tgf$.
The $x$-axis of Fig.~\ref{fig:tau gwtc-3} stops at $\tgf=0.0365$, as this is the largest possible value allowed when requiring that $p(\mu|z)$ asymptotes to $\mu^{-2}$ at large $\mu$.
    
Even for the most extreme optical depths, the probability of observing zero outliers is $P(k=0| \tgf) \geq$0.824
.
This translates into a very weak constraint on $\tgf$. 

The primary reason for this weak constraint is that the expected rate of outliers due to magnification is low.
Outliers are defined based on their \textit{observed} parameters, whereas the underlying population is defined in terms of true parameters.
The true and observed parameters can substantially differ in the presence of large measurement uncertainty, which are typical in GWTC-3 and for high-mass events in general~\cite{vitale_parameter_2017,fishbach_most_2020}.
An offset therefore exists between the high-mass truncation of the underlying mass distribution, $\mmax$, and the maximum mass needed to be considered an outlier, $m_{\max}^{\rm thresh}$.
This can be seen in the separation between the orange line and green region in Fig.~\ref{fig:schematic}.
For gravitational lensing to induce an outlier, it must supply a magnification large enough to overcome extremal measurement uncertainties, as typified by the 99th percentile of maximum observed masses.
Even with these magnifications, the system must have been intrinsically high-mass to become an outlier, and these are comparably rare: the \ac{BBH} mass distribution declines steeply above $\sim35\Msun$.
As we show in Appendix~\ref{ap:robustness checks:Lambda}, a less steep mass distribution would increase the expected number of high-mass outliers, but such a mass distribution is unlikely given the observed \ac{GW} data.
This means that, for typical events, substantial magnifications ($\mu \gtrsim 20$) are needed to produce outliers due to lensing alone. 
These magnifications are rare, even for the most extreme possible magnification distributions.

In sum, the lack of high-mass outliers in GWTC-3 is both unsurprising and weakly sensitive to the optical depth because lensing-induced outliers are rare events for all physical magnification distributions.

\section{Projections for future GW detectors}
\label{sec:results-future}

\begin{figure*}
    \centering
    \includegraphics[width=\linewidth]{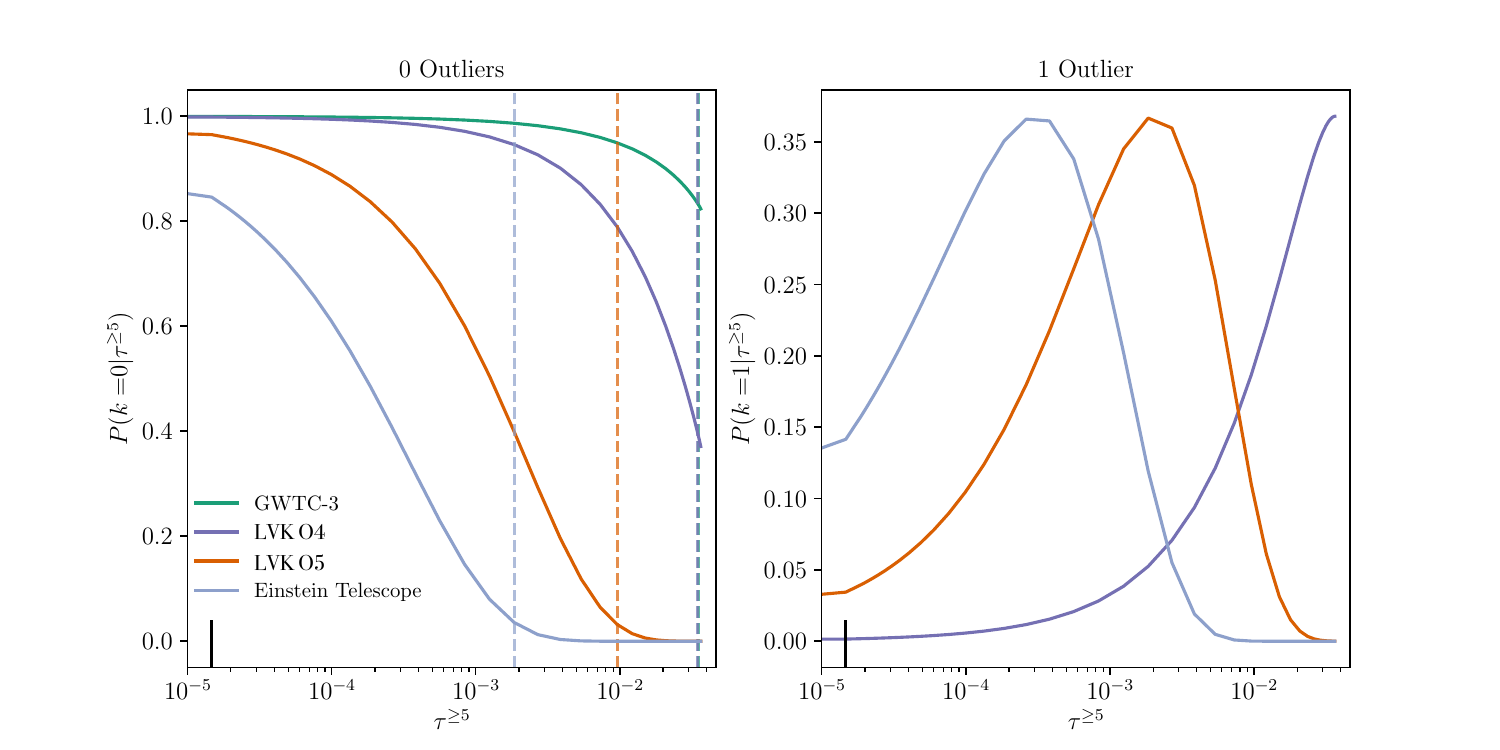}
    \caption{Probability of detecting zero (\emph{left panel}) and one (\emph{right panel}) high-mass outliers in current and future observing runs, as a function of $\tgf$.
    On the left panel, 95\% upper limits on $\tgf$ for each observing run are shown by vertical dashed lines.
    While increased observing time with no outlier detections will improve the upper limits on $\tgf$, these limits are orders of magnitude away from those expected for our Universe (\emph{black tickmark on both $x$-axes}).
    If one high-mass outlier is detected, weak constraints (rather than upper limits) on $\tau$ are possible.
    Note the different $y$-axis ranges between the two panels.
    }
    \label{fig:projections}
\end{figure*}

Having analyzed the currently-available \ac{GW} data, we study the future possibility of identifying lensing-induced outliers to the \ac{CBC} population.
We consider two scenarios -- one in which no outlier is identified and one in which a single outlier is observed -- and find that constraints on the optical depth will remain weak.
This again points to population outliers' lack of sensitivity to gravitational lensing. 

\subsection{Continued non-detection of outliers}
The probability of detecting zero high-mass \ac{BBH} outliers in current and future observing runs is plotted as a function of $\tgf$ in the left panel of Fig.~\ref{fig:projections}. 
The prediction for each observing run includes all previous observing runs (i.e. $N_{\exp}^{\rm outlier}(\tgf)|_{\rm O4} = N_{\exp}^{\rm outlier}(\tgf)|_{\rm O3} + N_{\exp}^{\rm outlier}(\tgf)|_{\rm O4\ only}$).
Transforming these probabilities into posteriors on $\tgf$ using Bayes' theorem results in $95\%$ upper-limits on $\tau$ shown by vertical dashed lines.
These limits are unsurprisingly weak: Fig.~\ref{fig:projections} suggests that it is unlikely to observe a high-mass outlier due to lensing in \ac{O4}, \ac{O5}, or one year of Einstein Telescope\footnote{While we only include explicit calculations for Einstein Telescope, we expect similar results to hold for other next-generation detectors, such as Cosmic Explorer, as their selection functions will be similar \cite{et_steering_committee_einstein_2020,evans_horizon_2021}.} observing, unless $\tgf>$0.035
, $\tgf>$0.010
, or $\tgf>$0.002
 respectively.
These limits are several orders of magnitude higher than values of $\tgf$ expected from cosmological simulations (black tickmark on the $x$-axes of Fig.~\ref{fig:projections}).

\begin{figure}
    \centering
    \includegraphics[width=\columnwidth]{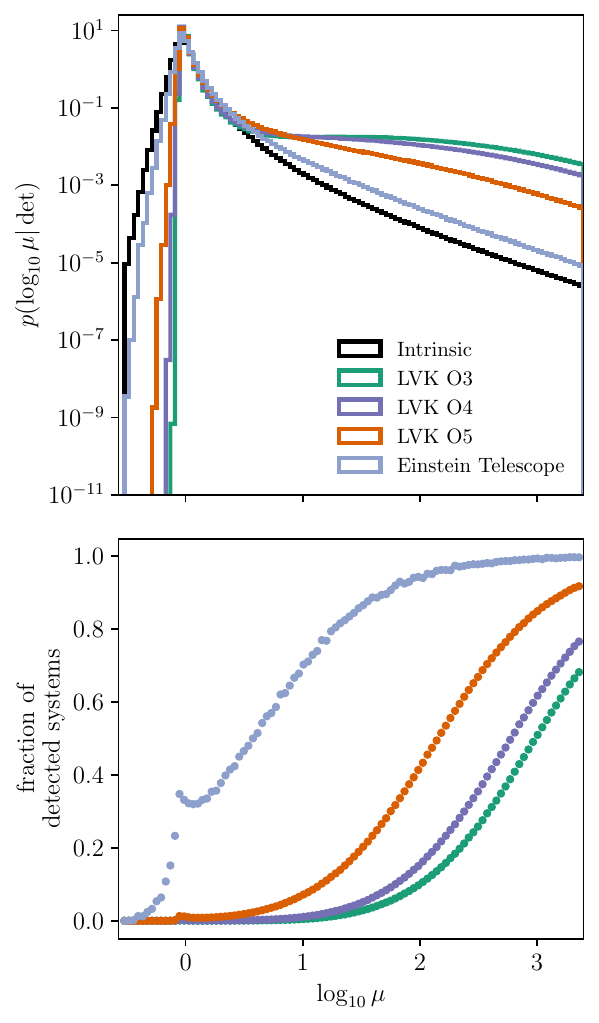}
    \caption{Detectability of events as a function of the magnification for current and future \ac{GW} detectors. 
    \emph{Top panel:} detected magnification distributions, assuming our fiducial $p(\mu|z)$ (black line in Fig.~\ref{fig:p mu tau}) and the \ac{BBH} population described in Section~\ref{sec:methods-astro pop}.
    \emph{Bottom panel:} the fraction of \acp{BBH} that are detected as a function of their magnification.
    The bottom panel is independent of the assumed magnification distribution, but it is impacted by the distribution of other BBH parameters.
    The current \ac{LVK} detectors and their planned upgrades have similar sensitivity to $\mu$ at the high magnifications required to become outliers, and nearly all \acp{BBH} with $\mu\gtrsim 1,000$ are observable by current detectors.
    Einstein Telescope's sensitivity will already be near-perfect so it is less impacted by magnification than the LVK detectors. 
    }
    \label{fig:Pdet mu}
\end{figure}

The improved constraint in \ac{O5} over \ac{O4} is due more to increased observing time than it is to detector sensitivity.
Integrating Eq.~(\ref{eq:Nexp MC}) over $t$ shows that $N_{\exp}^{\rm outlier}(\tgf)$ depends linearly on observing time, whereas increased detector sensitivity has a comparably smaller effect on $P_{\det + \mathrm{outlier}}$, and therefore on $N_{\exp}^{\rm outlier}(\tgf)$.
This latter point is demonstrated in Fig.~\ref{fig:Pdet mu}, where the top panel shows the expected distribution of detected magnifications for each detector considered, given the magnification and \ac{BBH} parameter distributions described in Section~\ref{sec:methods-astro pop}, and the bottom panel shows the probability of detecting a \ac{BBH} given its magnification.
Current detectors have similar detected magnification distributions, especially at the large magnifications ($\mu\gtrsim20$) needed to induce high-mass outliers. 
Thus, \emph{most lensing-induced outliers are already observable}, as their magnifications are large enough to bring them within the horizons of current detectors even when their true distances are much larger.
        
Still, lensing-induced outliers are rare enough that O4-like detectors would need to observe for $\sim1,000$ years to detect such an event if $\tgf=$$1.47\times 10^{-5}$, and 44
 years if $\tgf=$$1.24\times 10^{-3}$
. 

\subsection{Detection of one outlier}

Although it is unlikely that the \ac{LVK} will observe a lensing-induced outlier for expected values of the optical depth, it would be possible if $\tgf\gtrsim 0.026$. 
At such large optical depths, the rate of lensing would be high enough to substantially affect the inferred population, and the assumption made in Section~\ref{sec:methods-ffh} that our population was inferred with an unlensed sample may no longer be valid.
Nonetheless, we explore this scenario and find that it would result in the constraints on $\tgf$ shown in the right panel of Fig.~\ref{fig:projections}.
A detection of one high-mass outlier in \ac{O4} would mean that $\tgf\geq 7.3\times10^{-3}$, and the same detection in \ac{O5} or one year of Einstein Telescope observing would mean that $\tgf = 0.0039^{+0.011}_{-0.003}$ or $0.00084^{+0.00303}_{-0.00078}$ (median and 90\% credible intervals), respectively.
While it is promising that measurements of $\tgf$ are possible, these constraints are broad, spanning $2$--$3$ orders of magnitude.   

It is possible to improve the precision of these constraints by incorporating the additional information afforded by that outlier's apparent masses and distance.
For example, it is straightforward to calculate the magnifications that would make the outlier consistent with the population.
Such an approach was taken in \citet{abbott_properties_2020}, but with a theoretical population rather than the population inferred from the rest of the \ac{GW} data.
We explore this approach by simulating data that would arise from a large enough optical depth to observe one lensing-induced outlier in \ac{O4}.
We find that the lensing-induced outliers in this scenario typically have $\mu\gtrsim20$, while the magnifications required to make them consistent with the population are $\lesssim10$. 
Therefore, one would systematically underestimate the true magnification of a high-mass outlier.
This is a case of Eddington bias: the true source frame mass of the outlier is almost always less than its observed source frame mass, even without magnification, because events that are able to become outliers are scattered to higher observed masses.
This can be seen in Fig.~\ref{fig:schematic}, where the majority of squares in the right panel (i.e. events that will become outliers after lensing) are to the right of the orange curve (the maximum allowed true mass of the population) because measurement uncertainty has scattered them to higher detector-frame masses and/or lower distances.
In this way, measurement uncertainty mimics magnification.
Large magnifications are required to overcome the full effects of measurement uncertainty and cause outliers, but relatively smaller magnifications are needed to reach the upper edge of the measurement uncertainty distribution.
The mismatch between the true magnifications of outliers and the magnifications required for them to be consistent with the population is therefore a bias introduced by the failure to take measurement uncertainty into account.
Thus, attempts to measure the lensing magnification of a high-mass outlier and incorporate it into optical depth measurements will introduce systematic bias.

Additionally, the fact that $p(\mu|z)$ decays rapidly with increasing $\mu$ means that the detection of one event with  a magnification large enough to produce an outlier necessitates the existence of \emph{several other} events in the catalog with moderate magnifications.
The method presented in this work leaves out the information from these other magnified events, which may have moderate or low true masses and therefore not show up as high-mass outliers, but which would affect the inferred population.
This again points to an inconsistency in the method of using outliers from a population that is assumed to be inferred without the impacts of lensing. 
A self-consistent approach would be to simultaneously infer the \ac{CBC} population, the magnification distribution, and the magnifications received by individual events.
This is achieved with a hierarchical Bayesian analysis, and future work will develop this method and apply it to current and future observing runs.

\subsection{Identifying lensed BNSs in the lower mass gap}
We additionally explore the possibility of identifying lensed \acp{BNS} as population outliers. 
This is motivated by \citet{smith_discovering_2023}, who propose that the purported lower mass gap between the maximum neutron star mass and potential minimum black hole mass can aid in efficiently triggering electromagnetic follow-up of lensed signals.
If the true mass distribution vanishes between $[2.5\,\Msun,5\,\Msun]$ -- which is the most optimistic assumption for the prospect of identifying lensed signals in this region but is not required by the GW data \citep{farah_bridging_2022,abbott_population_2023} -- we find that the vast majority of systems in this mass range are not lensed. 
In O4, for example, we find that the expected number of events in the range $[2.5\,\Msun,5\,\Msun]$ will be 0.42 without the effects of gravitational lensing and 0.41 under our fiducial lens model.
These two numbers -- the false positive rate and the combined false and true positive rate, respectively -- are consistent with one another.
We therefore conclude that false positives dominate the lower mass gap region.
Furthermore, we apply the procedure described in Section~\ref{sec:methods-ffh} to the BNS population described in \citet{smith_discovering_2023} and find that an events' observed primary mass must be greater than $3.5\,\Msun$ to be robustly identified as an outlier.
A similar calculation can be done with the minimum black hole mass, showing that if the lower mass gap exists, it will be significantly polluted with events originating on either side of it, making it too narrow to be used for reliable identification of lensed \acp{BNS}.

\section{Conclusions}
\label{sec:discussion}
\acp{GW} that are magnified by gravitational lensing appear to originate from closer and more massive systems than would be inferred from their un-lensed signals.
This means that more high-mass population outliers should exist if the rate of gravitational lensing is high.
We explore the prospects for using the number of observed population outliers to constrain the rate of gravitational lensing. 
We use a definition of outliers that is informed by the astrophysical population inferred from existing GW data and accounts for detector noise fluctuations. 
This differs from previous work that tested the strong lensing hypothesis using astrophysical priors on the edges of the  mass distribution and only considered true \ac{CBC} masses, which noticeably differ from the noise-scattered observed masses in the parameter space of interest to lensing \citep{abbott_properties_2020,broadhurst_reinterpreting_2018,smith_discovering_2023}.
    
We find that for an event to appear as a population outlier, it must \emph{both} have a large magnification and a large true mass, both of which are rare. 
This has several consequences. 

First, it means that the detection of a lensing-induced outlier implies the existence of \emph{several} other highly-magnified, lower-mass events in the catalog.
Thus, an approach that simultaneously infers the \ac{CBC} population with the magnification distribution may be more sensitive to the optical depth, as it would use all of the available \ac{GW} data, not just the intrinsically high-mass systems.
        
Second, the number of observed high-mass outliers is only weakly sensitive to the high-magnification optical depth.
Even the most extreme possible optical depths are consistent with having observed zero outliers in GWTC-3.
Even when current detectors reach their design sensitivity, a lack of high-mass outliers will lead to an upper limit on the optical depth of 0.010
 at redshift 1, but this limit will still be $\sim3$ orders of magnitude above values of the optical depth expected from cosmological simulations, $\tgf\sim 10^{-5}$.
Increased detector sensitivity does not substantially improve the constraining power of this method, but significantly increased observing time does.
However, if a high-mass outlier is observed in future observing runs, a weak measurement of the strong lensing optical depth will be possible.

Third, identifying uncharacteristically high-mass systems may not be an effective method to determine which events are lensed given the rarity of lensing-induced outliers.
For example, while \citet{smith_discovering_2023} propose to use events in the lower mass gap as candidates for lensed BNS, non-lensed events make up the majority of detected signals with observed masses in that range. 
This high prevalence of false positives combined with the low expected number of events will make it difficult to identify single events as gravitationally lensed based on their uncharacteristically large masses alone. 
However, additional lensing information could complement this method. 
In particular, highly magnified events are expected to have similarly magnified copies with short time delays and known phase shifts \cite{lo_observational_2025,maria_ezquiaga_diffraction_2025}.

A few caveats should be considered with the results presented here.
Firstly, in the constraints on the optical depth reported in Sections~\ref{sec:results-gwtc3} and \ref{sec:results-future}, we assumed that all outliers are due to lensing.
However, false positives are possible, and non-magnified events cause about half of the observed high-mass outliers if the optical depth is small ($\approx10^{-5}$) and $\lesssim10\%$ of the observed outliers if the optical depth is large ($\gtrsim10^{-3})$.

Additionally, we have not taken into account the possibility that strongly lensed GWs could also be lensed by small-scale lenses such as stars or compact objects. 
This becomes more probable as the magnification of the event increases \cite{mishra_gravitational_2021}. 
In the case of lensing by small compact objects, frequency-dependent lensing effects could distort the waveform as much to prevent its detection with standard non-lensed template banks and match-filtered techniques \cite{chan_detectability_2025}. 
When distortions are milder but still measurable, they can be used to identify the event as lensed.
The outlier-based method described here could then serve as additional evidence for lensing.


Future work will simultaneously infer the \ac{GW} source population, the magnification distribution, and the magnifications of individual GW events.
This may be more effective at constraining the lensing optical depth than only considering high-mass outliers, which, as we have demonstrated, is relatively insensitive to the magnification distribution.
Inferring the high-magnification optical depth using \acp{GW} alone is a promising astrophysical and cosmological probe.
\ac{GW} catalogs are complete to high redshift, monitor the whole sky, and have known selection biases, making \acp{GW} complementary to galaxy survey-based probes of the magnification distribution and potentially more competitive than other transients. 
Moreover, the exquisite time resolution ($\sim$ msec) and long observing runs ($\sim$ years) of \ac{GW} detectors make \acp{GW} sensitive to essentially the entire range of lens masses, from stars to galaxy clusters. 
This implies that \ac{GW} observations can constrain the magnification distribution for \emph{all} lenses, which can be heavily affected by substructures \cite[e.g.][]{vujeva_effects_2025}. 
Altogether, including the effect of lensing in population analyses will \acp{GW} to probe the furthest compact binaries and discover new populations of compact objects.

\begin{acknowledgments}
We are grateful to Rico Lo, Matthew Mould, Aditya Vijaykumar and Luka Vujeva for helpful conversations.
A.M.F. is supported by the National Science Foundation Graduate Research Fellowship Program under Grant No. DGE-1746045.
This research was undertaken thanks in part to funding from the Canada First Research Excellence Fund through the Arthur B. McDonald Canadian Astroparticle Physics Research Institute.
JME is supported by the European Union’s Horizon 2020 research and innovation program under the Marie Sklodowska-Curie grant agreement No. 847523 INTERACTIONS, and by VILLUM FONDEN (grant no. 53101 and 37766). 
MF is supported by the  Natural Sciences and Engineering Research Council of Canada under Grant No. RGPIN-2023-05511, the University of Toronto Connaught Fund, the Alfred P. Sloan Foundation, and the Ontario Early Researcher Award. 
The Center of Gravity is a Center of Excellence funded by the Danish National Research Foundation under grant No. 184. 
The Tycho supercomputer hosted at the SCIENCE HPC center at the University of Copenhagen was used for supporting this work. 
This material is based upon work supported by NSF's LIGO Laboratory which is a major facility fully funded by the National Science Foundation.

\end{acknowledgments}

\appendix
\section{Tests of modeling assumptions}
\label{ap:robustness checks}
\subsection{Magnification model}
\label{ap:robustness checks:p_mu}
To determine the robustness of our results to our choice of magnification model, we calculate $N_{\text{exp}}^{\rm outlier}$ under an independently-calculated magnification model. 
We follow the prescription in \citet{oguri_effect_2018} to construct a magnification distribution that combines the effects of both weak and strong lensing.
For the weak lensing contribution, we calculate the probability density of convergences following the halo model of \citet{takahashi_probability_2011} and use `class` \cite{blas_cosmic_2011} to obtain a non-linear power spectrum.
For the strong lensing tail, we calculate strong-lensing statistics using cosmological simulations.
Under this model, we obtain an $N_{\text{exp}}^{\rm outlier} = 0.000339$.
Our fiducial model (Equation~\ref{eq:Dai17}, \citet{dai_effect_2017}) yields $N_{\text{exp}}^{\rm outlier} = 0.000170$.
These two numbers are similarly small, and their difference is less than the differences induced by Monte-Carlo uncertainty in our simulations.
The insensitivity of our results to the exact parameterization of our lens model is expected, as our results mostly depend on the high-magnification tail of the distribution, which is universal in shape across all lens models.

\subsection{Population model hyperparameters}
\label{ap:robustness checks:Lambda}
\begin{figure*}
    \includegraphics[width=\textwidth]{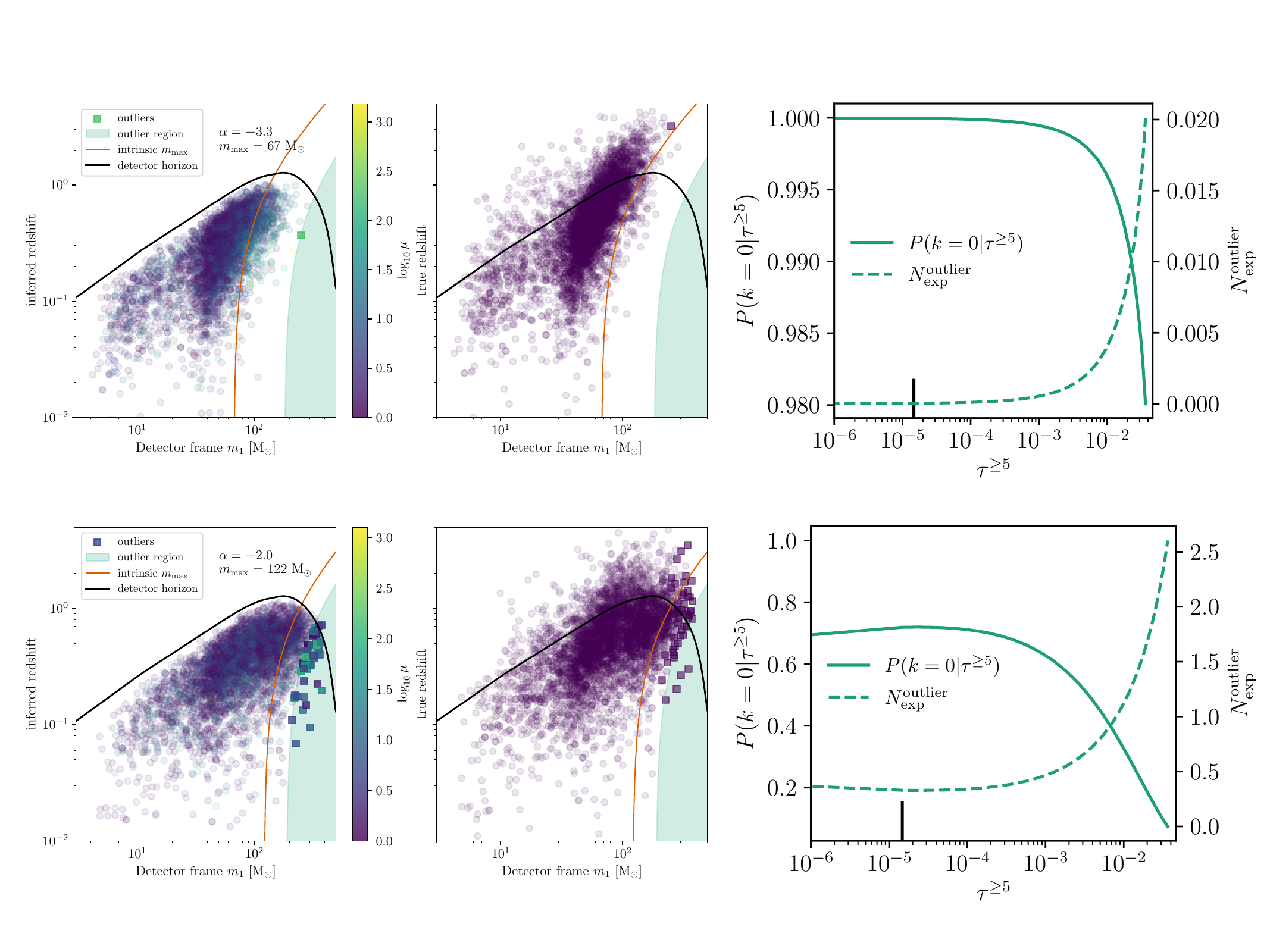}
    \label{fig:hyperparam extremes 1}
    \caption{Effect of varying the hyperparameters of the simulated population while keeping the outlier definition constant. The two leftmost columns should be compared to Fig.~\ref{fig:schematic} whereas the right column is similar to Fig.~\ref{fig:tau gwtc-3}.}
\end{figure*}
\begin{figure*}
    \includegraphics[width=\textwidth]{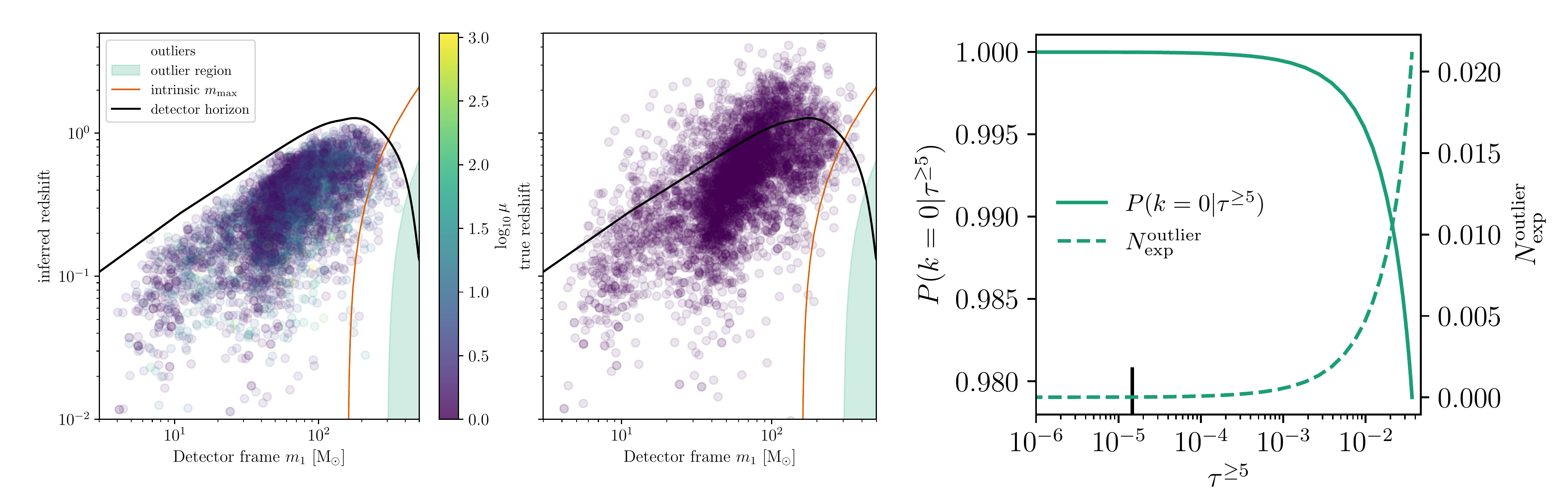}
    \label{fig:hyperparam extremes 2}
    \caption{Effect of using a large maximum mass ($160\Msun$) for both the outlier definition and simulated population.}
\end{figure*}

Our results are sensitive to the form of the \ac{BBH} population, as the rarity of intrinsically high-mass events contributes to the difficulty of producing lensing-induced, high-mass outliers.
This has motivated us to chose a model consistent with the \ac{GW} data.
Still, in this Section we perform two tests to determine how sensitive our conclusions are to the exact hyperparameters in our model.

In the first test, we construct the \ac{PPD} as in Section~\ref{sec:methods-ffh}, using the full hyperposterior obtained using a fit to GWTC-3.
We then calculate $N_{\text{exp}}^{\rm outlier}$ using  hyperparameter values that will yield the highest or lowest number of high-mass events.
This amounts to choosing the highest (lowest) value of the \ac{BBH} maximum mass and shallowest (steepest) value of the power law slope allowed by the hyperposterior.
These are very low-probability draws from the hyperposterior, as high maximum masses are strongly correlated with steep power law slopes.
We choose these values not as representative samples, but to maximize the effect of different hyperposterior draws.
This is in contrast to the procedure used in the main text and described in Section~\ref{sec:methods-Nexp}, which chooses the hyperposterior mean for all hyperparameters.

The results for both extremes are shown in Fig.~\ref{fig:hyperparam extremes 1}.
The top row presents the case where the rate of high-mass events is minimized.
As expected, fewer outliers are observed in the left two panels compared to Fig.~\ref{fig:schematic}, and no false positives are present.
This results in a minimal difference between the top right panel and \ref{fig:tau gwtc-3}, as both scenarios result in very few high-mass outliers and poor constraints on $\tgf$.
In the bottom row of Fig.~\ref{fig:hyperparam extremes 1}, we show the case where the rate of high-mass events is maximized.
Because of the mismatch between the population used to construct the outlier region and that used to simulate GW detections, the number of high-mass outliers is larger than that shown in Fig.~\ref{fig:schematic}, resulting in a higher number of expected outliers than that shown in Fig.~\ref{fig:tau gwtc-3}, and a constraint on the optical depth of $\tgf\leq0.0282$.
Thus, if the true \ac{BBH} mass distribution is far from our posterior mean, the number of high-mass outliers might be able to constrain the optical depth, but this constraint is still four orders of magnitude above its expected value.
This ability appears to be driven in part by the increased number of false positives, as shown by the prevalence of squares in the green region in the bottom-middle panel, as well as the fact that the solid green line does not intersect the $y$-axis at 1 in the bottom right panel.
Still, more true positives are present in this case than in our main analysis, demonstrating that the rarity of intrinsically high-mass events in the \ac{BBH} mass distribution significantly contributes to the small number of expected lensing-induced outliers.
Even in this extreme case, though, lensing does not appear to increase the number of expected high-mass outliers unless $\tgf\gtrsim10^{-3}$, as shown by the dashed line in the bottom-right panel of Fig.~\ref{fig:hyperparam extremes 1}.
Our conclusion that the number of high-mass outliers is a weak probe of the strong-lensing optical depth therefore remains the same, regardless of whether the true mass distribution is an unlikely draw from the inferred population.

The second test considers hyperparameters that are not consistent with GWTC-3, but might be able to accommodate the newly-detected high-mass event, GW231123~\citep{the_ligo_scientific_collaboration_gw231123_2025}. 
We choose a maximum mass of $160\Msun$ and take the hyperposterior mean for all other parameters, and use this choice to calculate both the \ac{PPD} and $N_{\text{exp}}^{\rm outlier}$.
The results are shown in Fig.~\ref{fig:hyperparam extremes 2}.
Interestingly, even fewer outliers are detectable under this scenario than the one considered in the main text.
This may point to the larger maximum mass producing a  more stringent outlier definition that simulated data is not able to overcome. 
Regardless, the differences between the rightmost panel of Fig.~\ref{fig:hyperparam extremes 2} and Fig. \ref{fig:tau gwtc-3} is minimal.

As a result of these two tests, we determine that our conclusions are not extremely sensitive to the exact details of the high-mass \acp{BBH} distribution's shape, despite their dependence on the fact that high-mass \acp{BBH} are rare.

\subsection{Statistical threshold defining an outlier}
\label{ap:robustness checks:P_thresh}

\begin{figure*}
    \includegraphics[width=\textwidth]{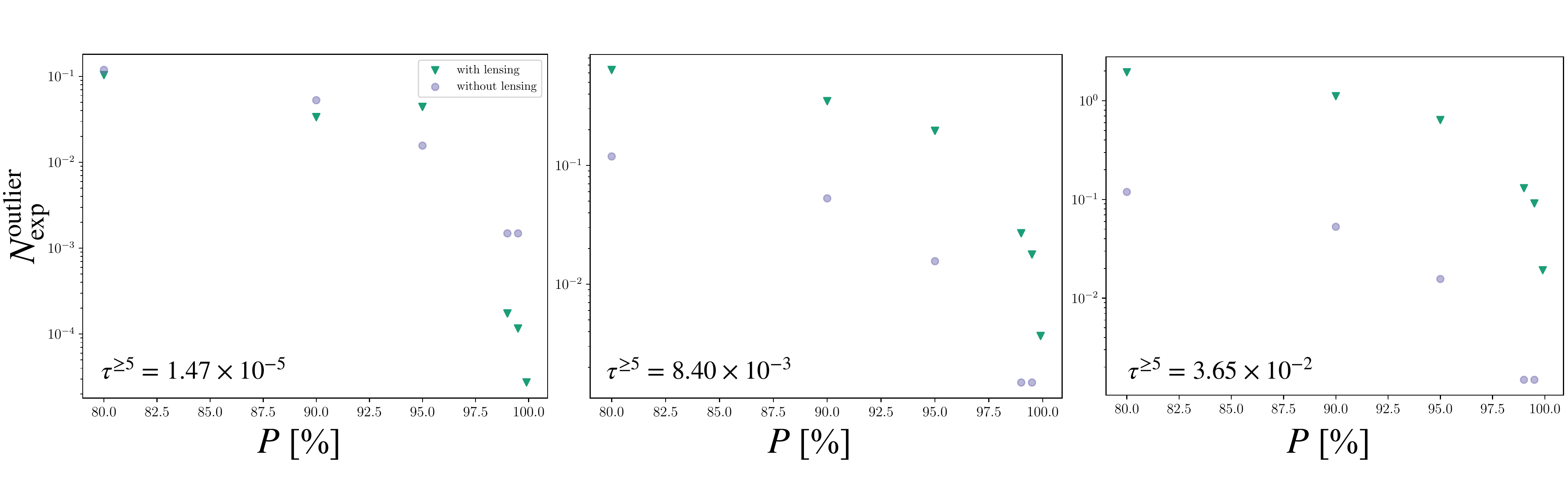}
    \label{fig:Nexp vs P}
    \caption{Expected number of high-mass outliers as a function of $P$ (defined in Section~\ref{sec:methods-ffh}). Green triangles include true and false positives, as they include the effects of both gravitational lensing and statistical fluctuations. Violet circles include only false positives, as they only include the effects of statistical fluctuations. For all optical depths considered, the number of expected true and false positives decreases with more stringent outlier definitions. Note the differing $y$-axis ranges in each panel.}
\end{figure*}

\begin{figure}
    \centering
\includegraphics[width=\linewidth]{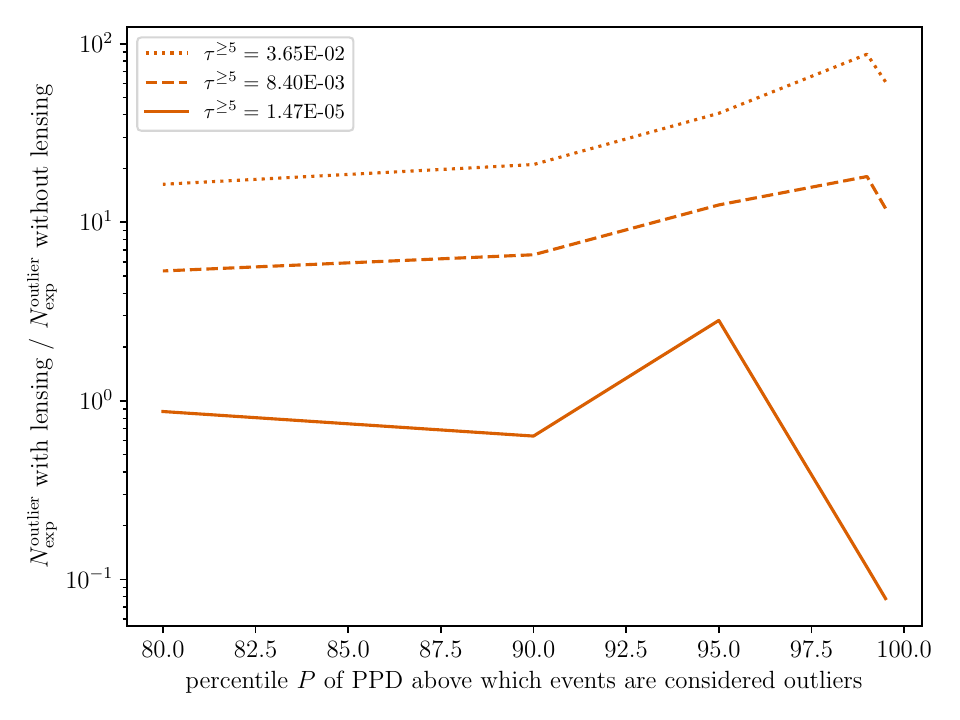}
    \caption{Ratio of the number of both of true and false positives to the number of false positives, as a function of $P$ for three different optical depths. This is a single simulation, and the exact $y$-axis values may change with different iterations. In general, more stringent thresholds lead to fewer false positives compared to true positives, but the exact value of $P$ that optimizes this ratio varies depending on the optical depth considered.}
    \label{fig:true_to_false}
\end{figure}

In this Section, we examine how our results are affected by the choice of percentile, $P$, defining which events are identified as outliers.
The green triangles in Fig.~\ref{fig:Nexp vs P} show the expected number of high-mass outliers as a function of $P$ for three different optical depths.
Higher values of $P$ lead to fewer expected outlier detections for all optical depths considered.
While this may seem to imply that lowering $P$ would lead to a higher sensitivity to gravitational lensing, $N_{\rm exp}^{\rm outlier}$ includes outliers due to lensing and those due to statistical fluctuations, both of which are affected by the choice of $P$.
Thus, it is important to consider the effect of $P$ on the number of ``false positives'', i.e. events that are identified as outliers because of statistical fluctuations rather than gravitational lensing.
To determine the number of false positives, we repeated simulations described in Section~\ref{sec:methods-Nexp}, but did not add the effects of gravitational lensing. 
In other words, all events were assigned $\mu=1$. 
The expected number of false positives is plotted with violet points in Fig.~\ref{fig:Nexp vs P}, and also decreases as the threshold becomes more stringent.

Ideally, $P$ would be chosen such that the ratio of false to true positives is minimized.
As shown in Fig~\ref{fig:true_to_false}, we find that the value of $P$ that minimizes this ratio depends on the optical depth.
The peak of the ratio of true to false positives is lower (i.e. less stringent) for low optical depths and higher for high optical depths. 
Thus, there is no universally optimal choice for $P$ and we arbitrarily choose $P=99$. 

\bibliography{references}
\end{document}